\documentclass[letterpaper, 10 pt, conference]{ieeeconf}

\IEEEoverridecommandlockouts
\usepackage{amsmath,amsfonts,amssymb}
\usepackage{graphics}

\newtheorem{theorem}{Theorem}[section]
\newtheorem{proposition}[theorem]{Proposition}

\newtheorem{remark}[theorem]{Remark}

\newtheorem{corollary}[theorem]{Corollary}

\newtheorem{algorithm}[theorem]{Algorithm}

\newcommand{\Hi}{{\cal H}_\infty}
\newcommand{\Lxi}{\mathcal{L}_{\xi}}
\newcommand{\LxiN}{\mathcal{L}_{\xi}^N}
\newcommand{\R}{\mathbb{R}}
\newcommand{\C}{\mathbb{C}}

\title{\LARGE \bf Computing $\Hi$ Norms of Time-Delay Systems}

\author{\IEEEauthorblockN{Suat Gumussoy and Wim Michiels}
\IEEEauthorblockA{Department of Computer Science, K. U. Leuven\\
Celestijnenlaan 200A, 3001, Heverlee, Belgium\\ Emails:
suat.gumussoy@cs.kuleuven.be, wim.michiels@cs.kuleuven.be}}

\author{Suat Gumussoy and Wim Michiels
\thanks{S. Gumussoy and W. Michiels are with Department of Computer Science, K. U. Leuven
        Celestijnenlaan 200A, 3001, Heverlee, Belgium\
        {\tt\small Emails: suat.gumussoy@cs.kuleuven.be, wim.michiels@cs.kuleuven.be}}%
}

\begin{document}

\maketitle
\thispagestyle{empty}
\pagestyle{empty}

\begin{abstract}
In this paper we consider the computation of $\Hi$ norm of retarded time-delay systems with discrete pointwise state delays. It is well known that in the finite dimensional case $\Hi$ norm of a system is computed using the connection between the singular values of the transfer function and the imaginary axis eigenvalues of an Hamiltonian matrix. We show a similar connection between the singular values of a transfer function of a time-delay system and the imaginary axis eigenvalues of an infinite dimensional operator $\Lxi$. Using spectral methods, this linear operator is approximated with a matrix. The approximate $\Hi$ norm of the time-delay system is calculated using the connection between the imaginary eigenvalues of this matrix and the singular values of a finite dimensional approximation of the time-delay system. Finally the approximate results are corrected by solving a set of equations which are obtained from the reformulation of the eigenvalue problem for $\Lxi$ as a finite dimensional nonlinear eigenvalue problem.
\end{abstract}

\section{Introduction}
In robust control of linear systems,  stability and performance criteria are often expressed by $\Hi$ norms of appropriately defined transfer functions. Therefore, the availability of robust methods to compute $\Hi$ norms is essential in a computer aided control system design \cite{zhou}.

The computation of $\Hi$ norm for the finite dimensional plants is based on the relation between the existence of the singular values of the transfer function equal to the fixed value and the existence of the imaginary axis eigenvalues of the corresponding Hamiltonian matrix of the same fixed value \cite{bisection:byers}. This relation allows the computation of $\Hi$ norm via the well-known level set method \cite{bisection:boyd}. It is possible to set the level for the singular values of the transfer function using the relation above and achieve quadratically convergent algorithms in $\Hi$ norm computation for finite dimensional plants \cite{quad:boyd, quad:bruinsma}.

In this paper, we consider the computation of the
$\mathcal{H}_{\infty}$ norm of the stable time-delay system $G$ with the transfer function representation,
\vspace{-.2cm}
\begin{equation}\label{eq:tfG}
{\textstyle G(s)=C\left(s I-A_0-\sum_{i=1}^m A_i
e^{-\tau_i s}\right)^{-1}B+ D}
\vspace{-.2cm}
\end{equation}
where the system matrices are $A_i^{n\times n}$, $B^{n\times n_u}$,
$C^{n_y\times n}$, $D^{n_y\times n_u}$, $\;i=0,\ldots,m$ are
real-valued and the time delays, $(\tau_1,\ldots,\tau_m)$, are nonnegative real
numbers. Equivalently, the $\Hi$ norm of (\ref{eq:tfG}) is defined as the largest singular value of the $G(j\omega)$ over all the frequency interval.

In Section \ref{sec:Lxi}, it is shown that given $\xi>0$, the existence of the singular values of the transfer function (\ref{eq:tfG}) equal to $\xi$ is equivalent to the existence of the imaginary axis eigenvalues of the linear infinite-dimensional operator $\Lxi$.

By this relation, we extended the level set methods to the time-delay systems. The difference lies in the fact that in every iteration of the level $\xi$, the imaginary axis eigenvalues of the infinite-dimensional linear operator $\Lxi$ are required instead of that of Hamiltonian matrix in the finite dimensional delay-free case.

In Section \ref{sec:LxiN}, we approximate the infinite-dimensional operator $\Lxi$ by a finite-dimensional matrix approximation $\LxiN$. We show that for a fixed level set $\xi>0$, there is a relation between the imaginary axis eigenvalues of the matrix $\LxiN$ and the singular values of a finite-dimensional approximation of $G$ equal to $\xi$. Therefore, the $\Hi$ norm calculated by the level set methods and $\LxiN$ is the $\Hi$ norm of the finite dimensional approximation of $G$.

In Section \ref{sec:HinfCor}, we correct the approximate results by using the property that the eigenvalues of the linear infinite dimensional operator $\Lxi$ appear as solutions of a finite dimensional nonlinear eigenvalue problem. This allows to write the conditions to characterize the peaks in singular value plot and correct the approximate $\Hi$ norm.

Two numerical algorithms based on level set methods \cite{bisection:byers,quad:bruinsma} for $\Hi$ norm computation of the time-delay system are given in Section \ref{sec:Alg}. A numerical example and concluding remarks are given in Section \ref{sec:Ex} and \ref{sec:Conc}.

\textbf{Notation:} \\
The notation in the paper is standard and given below.\vspace{.2cm}
\begin{tabular}{rl}
  $\C, \R:$ & the field of the complex and real numbers, \\
  $\mathbb{C}^{n}:$ & n-dimensional complex space, \\
  $A^{*}:$ & complex conjugate transpose of the matrix $A$, \\
  $A^{-T}:$ & transpose of the inverse matrix of $A$, \\
  $\mathcal{D}(.):$ & domain of an operator, \\
  $\sigma_i(A):$ & i$^\textrm{th}$ singular value of $A$, \\
  $\Re(u):$ & real part of the complex number $u$, \\
  $\Im(u):$ & imaginary part of the complex number $u$. \\
  $\det(A):$ & determinant of the matrix $A$. \\
  $\tau_{\max}:$ & the maximum of the delays $(\tau_1,\ldots,\tau_m)$ in (\ref{eq:tfG}). \\
  $\mathcal{C}:$ &  the space of continuous complex functions.
\end{tabular}

\section{Linear Infinite-Dimensional Eigenvalue Problem} \label{sec:Lxi}

The connection between the singular values of a transfer function and the
imaginary eigenvalues of a corresponding Hamiltonian matrix is
given in \cite{bisection:byers,quad:boyd} that laid the basis for
the established level set methods to compute $\mathcal{H}_{\infty}$
norms. The following theorem generalizes this connection to the
time-delay systems:

\begin{theorem} \label{thm:GLxi}
Let $\xi> 0$ be such that the matrix
\vspace{-.2cm}
\[
D_{\xi}:=D^TD-\xi^2 I
\vspace{-.2cm}
\]
is non-singular.
For $\omega\geq 0$, the matrix $G(j\omega)$ has a
singular value equal to $\xi>0$ if and only if $\lambda=j\omega$ is an eigenvalue of the linear infinite dimensional operator $\Lxi$ on $X:=\mathcal{C}([-\tau_{\max},\ \tau_{\max}],\mathbb{C}^{2n})$ which is defined by
\vspace{-.2cm}
\begin{multline}
\mathcal{D}(\mathcal{L}_{\xi}) =\left\{\phi\in X:
\phi^{\prime}\in X,\hspace{3.5cm} \right. \label{def:Lxi2} \\
\phi^{\prime}(0)=M_{0}\phi(0)  + \sum_{i=1}^m
(M_i\phi(-\tau_i)+M_{-i}\phi(\tau_i) ) \},
\end{multline}
\vspace{-.3cm}
\begin{equation}
{\textstyle \mathcal{L}_{\xi}\phi=\phi^{\prime},\;\phi\in\mathcal{D}(\Lxi)} \label{def:Lxi}
\end{equation}
\vspace{-.2cm}
with
\[{\textstyle
\begin{array}{l}
M_{0}=\left[\begin{array}{cc}  A_0-B D_{\xi}^{-1} D^T C& -B
D_{\xi}^{-1}B^T\\
\xi^2 C^T D_{\xi}^{-T} C & -A_0^T+C^TD D_{\xi}^{-1} B^T
\end{array}\right],\\
M_i=\left[\begin{array}{cc} A_i &0\\0&0
\end{array}\right],\ \
M_{-i}=\left[\begin{array}{cc} 0 &0\\0&-A_i^T
\end{array}\right],\ \ 1\leq i\leq N.
\end{array}}
\]
\end{theorem}
\noindent\textbf{Proof.\ } The proof is given in Appendix, Section \ref{sec:App}.

Equivalently Theorem \ref{thm:GLxi} can be stated that there is a singular value of $G$ equal to $\xi$ at $\omega=\omega_0$, $\sigma_i(G(j\omega_0))=\xi$, if and only if the eigenvalue problem for the linear operator $\Lxi$
\begin{equation}\label{prob:LxiEig}
{\textstyle (\lambda I-\mathcal{L}_{\xi}) u=0:\ \lambda\in\mathbb{C},\ u\in X,\
u\neq 0.}
\end{equation}
has a solution for $\lambda=j\omega_0$.

Although the operator $\Lxi$ generally has infinite number of eigenvalues, one can show that the number of eigenvalues on the imaginary axis is always finite. Therefore, eigenvalue problem (\ref{prob:LxiEig}) is computationally well-posed.

\begin{proposition}\label{prop:LxiSym}
$\lambda$ is an eigenvalue of the linear operator $\Lxi$ if and only if $-\bar{\lambda}$ is an eigenvalue of the linear operator $\Lxi$.
\end{proposition}
\noindent\textbf{Proof.\ } The proof is given in Appendix, Section \ref{sec:App}.

By Proposition \ref{prop:LxiSym}, the set of eigenvalues of $\Lxi$ is symmetric with respect to the imaginary axis. In the delay-free case, the operator $\Lxi$ reduces to a Hamiltonian matrix.

The key role of Theorem \ref{thm:GLxi} is that it reduces the $\Hi$ norm computation of (\ref{eq:tfG}) into the bisection search for maximum level set $\xi$ for which the linear operator $\Lxi$ has imaginary axis eigenvalues.

Instead of solving the difficult linear infinite dimensional eigenvalue problem, we can use the connection in Theorem~\ref{thm:GLxi} and apply the level set methods for the $\Hi$ norm computation of (\ref{eq:tfG}) in two steps:
\begin{enumerate}
  \item[1)] The approximate solution of the eigenvalue problem can be calculated by solving the standard linear eigenvalue problem of the discretized linear operator of $\Lxi$.
  \item[2)] The approximate results can be corrected by using the property that the eigenvalues of the linear infinite dimensional operator $\Lxi$ appear as solutions of a finite dimensional nonlinear eigenvalue problem.
\end{enumerate}

The approximation of the linear operator $\Lxi$ and the corresponding standard eigenvalue problem (\ref{prob:LxiEig}) is given in Section \ref{sec:LxiN}. The correction algorithm of the approximate results in the second step is explained in Section \ref{sec:HinfCor}.

\section{Finite-dimensional Approximation} \label{sec:LxiN}

In this section, the linear infinite dimensional eigenvalue problem (\ref{prob:LxiEig}) is discretized based on approximating the infinite-dimensional operator $\mathcal{L}_{\xi}$ by a matrix using a \emph{spectral method} (see, e.g. \cite{trefethenspectral,breda,bredaNonLocal}). Given a positive integer $N$, we consider a mesh $\Omega_N$ of $2
N+1$ distinct points in the interval $[-\tau_{\max},\ \tau_{\max}]$:
\begin{equation}\label{defmesh}
{\textstyle \Omega_N=\left\{\theta_{N,i},\ i=-N,\ldots,N\right\}},
\end{equation}
where
\begin{displaymath}
-\tau_{\max}\leq\theta_{N,-N}<\ldots<\theta_{N,0}=0<\cdots<\theta_{N,N}\leq\tau_{\max}.
\end{displaymath}

This allows to replace the continuous space $X$ with the space $X_N$
of discrete functions defined over the mesh $\Omega_N$, i.e.\ any
function $\phi\in X$ is discretized into a block vector
$x=[x_{-N}^T\cdots\ x_{N}^T ]^T\in X_N$ with components
\[{\textstyle
x_i=\phi(\theta_{N,i})\in\mathbb{C}^{2n},\ \ i=-N,\ldots,N}.
\]
Let $\mathcal{P}_N x,\ x\in X_N$ be the unique $\mathbb{C}^{2n}$
valued interpolating polynomial of degree $\leq 2N$ satisfying
\[{\textstyle
\mathcal{P}_N x (\theta_{N,i})=x_{i},\ \ i=-N,\ldots,N}.
\]
In this way, the operator $\mathcal{L}_{\xi}$ over $X$ can be
approximated with the matrix $\mathcal{L}_{\xi}^N:\ X_N\rightarrow
X_N$, defined as
\vspace{-.1cm}
\begin{eqnarray}
\hspace*{-1.4cm} \nonumber \left(\mathcal{L}_{\xi}^N\ x\right)_i&=&\left(\mathcal{P}_N
x\right)^{\prime}(\theta_{N,i}),\quad i=-N,\ldots,-1,1,\ldots,N \\
\nonumber \left(\mathcal{L}_{\xi}^N\ x\right)_0&=&M_0 \mathcal{P}_N x(0)+\sum_{i=1}^m (M_i \mathcal{P}_N x(-\tau_i) \\
& & \hspace{3cm}+ M_{-i} \mathcal{P}_N x(\tau_i)) \label{defldisc}
\end{eqnarray}

Using the Lagrange representation of $\mathcal{P}_N x$,
\[{\textstyle
\begin{array}{l}
\mathcal{P}_N x=\sum_{k=-N}^N l_{N,k}\ x_k,\
\end{array}}
\]
where the Lagrange polynomials $l_{N,k}$ are real valued polynomials
of degree $2N$ satisfying
\vspace{-.1cm}
\[{\textstyle
l_{N,k}(\theta_{N,i})=\left\{\begin{array}{ll}1 & i=k,\\
0 & i\neq k,
\end{array}\right.}
\]
 we obtain the explicit form\vspace{-.1cm}
\[{\scriptstyle
\mathcal{L}_{\xi}^N=
\left[\begin{array}{lll}
d_{-N,-N} &\hdots & d_{-N,N} \\
\vdots & & \vdots \\
d_{-1,-N} &\hdots & d_{-1,N} \\
a_{-N} & \hdots & a_N\\
d_{1,-N} &\hdots & d_{1,N} \\
\vdots & & \vdots \\
d_{N,-N} &\hdots & d_{N,N}
\end{array}\right]\in\mathbb{R}^{(2N+1)(2n)\times(2N+1)2n}},
\]
where\vspace{-.1cm}
\[{\textstyle
\begin{array}{lll}
d_{i,k}&=&l^{\prime}_{N,k}(\theta_{N,i}) I,\ \ \ \
i,k\in\{-N,\ldots,N\},\;i\neq0\\
a_0&=& M_0\ x_0+\sum_{k=1}^m \left(M_k
l_{N,k}(-\tau_k)+M_{-k}l_{N,k}(\tau_k)\right) \\
a_{k}&=&\sum_{k=1}^m \left(M_k
l_{N,k}(-\tau_k)+M_{-k}l_{N,k}(\tau_k)\right) \\
&& \hspace{-1.3cm}
k\in\{-N,\ldots,N\},\ k\neq0.
\end{array}}
\]
\vspace{-.1cm}
Note that all the problem specific information and the parameter $\xi$ are concentrated in the middle row of $\mathcal{L}_{\xi}^N$, i.e.\ the elements $(a_{-N},\ldots,a_N)$, while all other elements of $\mathcal{L}_{\xi}^N$ can be computed beforehand.

The matrix $\LxiN$ is a dense matrix with dimensions \mbox{$(2N+1)(2n)\times(2N+1)(2n)$}. Using the  approach at Section $2.2.2$ in \cite{KVerheyden} based on appropriate choice of the polynomial basis and the grid, the eigenvalue problem for $\LxiN$ can be written as a sparse generalized eigenvalue problem. Therefore, large-scale methods can be utilized for the linear eigenvalue problem.

Since the methods  for computing $\mathcal{H}_{\infty}$ norms proposed in \cite{genin} are based on checking the presence
of eigenvalues of $\LxiN$ on the imaginary axis and thus strongly rely on the symmetry of the eigenvalues with respect to the imaginary axis, it is important that this property is \emph{preserved} in the discretization. The following Proposition gives the condition on the mesh such that this symmetry holds.

\begin{proposition}\label{propsymLxi}
If the mesh $\Omega_{N}$ satisfies \vspace{-.1cm}
\begin{equation}\label{meshSymmetry}
\theta_{N,-i}=-\theta_{N,i},\ i=1,\ldots,N,
\vspace{-.2cm}
\end{equation}
then the following result hold:
for all $\lambda\in\mathbb{C}$, we have
\begin{equation}\label{symextraLxi}
\det\left(\lambda I - \LxiN\right)=0 \Leftrightarrow \det\left(-\bar{\lambda}-\LxiN\right)=0.
\end{equation}
\end{proposition}

\noindent\textbf{Proof.\ } The proof is given in Appendix, Section \ref{sec:App}.

We are primarily interested in the eigenvalues of $\Lxi$ on the imaginary axis. These eigenvalues  are typically among the smallest eigenvalues and one can easily show that the individual eigenvalues of $\LxiN$ exhibit spectral convergence
 to the corresponding eigenvalues of $\Lxi$ (following the lines of \cite{breda}). Since the symmetry property of the
spectrum is preserved in the discretization, a small value of $N$ is sufficient in most practical problems for computing a good
approximation of the $\Hi$-norm which can be employed as a starting point for a direct computation.

\section{Correction of $\Hi$ Norm} \label{sec:HinfCor}

By using the finite dimensional level set methods, the largest level set $\xi$ where $\LxiN$ has imaginary axis eigenvalues and their corresponding frequencies are computed. In the correction step, these approximate results are corrected by using the property that the eigenvalues of the $\Lxi$ appear as solutions of a finite dimensional nonlinear eigenvalue problem. The following theorem establishes the link between the linear infinite dimensional (\ref{prob:LxiEig}) and the nonlinear eigenvalue problem.

\begin{theorem} \label{thm:Lxi-Hxi}
Let $\xi> 0$ be such that the matrix
\vspace{-.2cm}
\[
D_{\xi}:=D^TD-\xi^2 I
\vspace{-.2cm}
\]
is non-singular. Then, $\lambda$ is an eigenvalue of linear operator $\Lxi$ if and only if
\vspace{-.2cm}
\begin{equation} \label{prob:HxiEigThm}
\det H_{\xi}(\lambda)=0,
\vspace{-.2cm}
\end{equation}
where
\vspace{-.3cm}
\begin{equation} \label{eq:HamMatrix}
H_{\xi}(\lambda):=\lambda I-M_0-\sum_{i=1}^m \left(M_i
e^{-\lambda\tau_i}+M_{-i}e^{\lambda\tau_i}\right)
\vspace{-.1cm}
\end{equation}
and the matrices $M_0$, $M_i$, $M_{-i}$ are defined in Theorem \ref{thm:GLxi}.
\end{theorem}

\noindent\textbf{Proof.\ } The proof is given in Appendix, Section \ref{sec:App}.

By Proposition \ref{prop:LxiSym} and Theorem \ref{thm:Lxi-Hxi}, the eigenvalues of the nonlinear eigenvalue problem (\ref{prob:HxiEig}) are symmetric with respect to the imaginary axis similar to the Hamiltonian matrix in the delay-free case.

The solutions of (\ref{prob:HxiEigThm}) can be
found by solving \vspace{-.1cm}
\begin{equation}
{\textstyle H_{\xi}(\lambda)\ v=0,\ \ \lambda\in\mathbb{C},\
v\in\mathbb{C}^{2n},\ v\neq 0}, \label{prob:HxiEig}
\end{equation}
which in general has an infinite number of solution.

Theorem \ref{thm:GLxi} and \ref{thm:Lxi-Hxi} establish the connections between the singular values of the transfer function of (\ref{eq:tfG}), the linear infinite dimensional eigenvalue problem (\ref{prob:LxiEig}), and the nonlinear eigenvalue problem (\ref{prob:HxiEig}).

The correction method is based on the property that if \mbox{$\hat{\xi}=\|G(j\omega)\|_{\mathcal{H}_{\infty}}$}, then (\ref{prob:HxiEig}) has a multiple non-semisimple eigenvalue:

If $\hat\xi\geq 0$ and $\hat\omega\geq 0$ are such that\vspace{-.2cm}
\begin{equation}\label{direct0}
\|G(j\omega)\|_{\mathcal{H}_{\infty}}=\hat\xi=\sigma_1(G(j\hat\omega)),
\vspace{-.2cm}
\end{equation}
then setting \vspace{-.2cm}
\[
h_{\xi}(\lambda)=\det H_{\xi}(\lambda),
\vspace{-.15cm}
\]
the pair $(\hat\omega,\hat\xi)$ satisfies\vspace{-.1cm}
\begin{equation}\label{direct1}
h_{\xi}(j\omega)=0,\ \ h_{\xi}^{\prime}(j\omega)=0.
\vspace{-.1cm}
\end{equation}
These complex-valued equations seem over-determined but this is not
the case due to the spectral properties of $\mathcal{H}_{\xi}(\lambda)$. Using the symmetry of the eigenvalues of the nonlinear eigenvalue problem (\ref{prob:HxiEig}) with respect to imaginary axis, we can write the following:
\begin{corollary}\label{corextra}
For $\omega\geq 0$, we have\vspace{-.2cm}
\begin{equation}\label{col1}
\Im\ h_{\xi}(j\omega)=0
\vspace{-.2cm}
\end{equation}
 and\vspace{-.1cm}
\begin{equation}\label{col2}
 \Re\ h_{\xi}^{\prime}(j\omega)=0.
\end{equation}
\end{corollary}
\noindent\textbf{Proof.\ } From the symmetry property of the eigenvalues with respect to the imaginary axis,\vspace{-.1cm}
\[
h_{\xi}(\lambda)=h_{\xi}(-\lambda),\ \ \ h_{\xi}^{\prime}(
\lambda)=-h_{\xi}^{\prime}( -\lambda).
\vspace{-.1cm}
\]
Substituting $\lambda=j\omega$ yields\vspace{-.1cm}
\[{\textstyle
\begin{array}{l}
h_{\xi}(j\omega)=h_{\xi}(-j\omega)=\left(h_{\xi}(j\omega)\right)^*, \\
h_{\xi}^{\prime}(j\omega)=-h_{\xi}^{\prime}(-j\omega)=-\left(h_{\xi}^{\prime}(j\omega)\right)^*,
\end{array}}
\vspace{-.2cm}
\]
and the assertions follow. \hfill $\Box$

\noindent Using Corollary~\ref{corextra} we can simplify the
conditions (\ref{direct1}) to:\vspace{-.2cm}
\begin{equation}\label{direct2x}
{\textstyle \left\{\begin{array}{l}
\Re\ h_{\xi}(j\omega)=0 \\
\Im\ h_{\xi}^{\prime}(j\omega)=0
\end{array}\right..}
\vspace{-.2cm}
\end{equation}
Hence, the pair $(\hat\omega,\hat\xi)$ satisfying (\ref{direct0})
can be directly computed from the two equations (\ref{direct2x}),
e.g.\ using Newton's method, provided that good starting values are
available.

\noindent The drawback of working directly with (\ref{direct2x}) is
that an explicit expression for the determinant of $H_{\xi}$ is
required. To avoid this, let $u,v\in\mathbb{C}^n$ be such that\vspace{-.2cm}
\[{\textstyle
H_{\xi}(j\omega) \left[\begin{array}{c}u\\
v\end{array}\right]=0,\ \ \ n(u,v)=0},
\vspace{-.1cm}
\]
where $n(u,v)=0$ is a normalizing condition. Given the structure of
$H_{\xi}$ it can be verified that a corresponding left eigenvector
is given by $[-v^*\ u^*]$. According to \cite{lancaster}, we get\vspace{-.2cm}
\[{\textstyle
h_{\xi}'(j\omega)=0\Leftrightarrow [-v^*\ u^*]\ H^{\prime}_{\xi}(j\omega) \left[\begin{array}{c}u\\
v\end{array}\right]=0}.
\vspace{-.2cm}
\]
A simple computation yields:\vspace{-.2cm}
\begin{equation}
{\scriptstyle
[-v^*\ u^*]\ H_{\xi}^{\prime}(j\omega) \left[\begin{array}{c}u\\
v\end{array}\right]=2\Im\left\{v^*\left(I+\sum_{i=1}^p A_i\tau_i
e^{-j\omega\tau_i}\right)u\right\}},
\end{equation}
which is always real. This is a consequence of the property
(\ref{col2}).

\noindent Taking into account the above results, we end up with
$4n+3$ real equations\vspace{-.2cm}
\begin{equation}\label{forfinal}
{\textstyle \left\{\begin{array}{l}
H(j\omega,\ \xi)\left[\begin{array}{c}u, \\
v\end{array}\right]=0, \quad n(u,v)=0\\
\Im\left\{v^*\left(I+\sum_{i=1}^p A_i\tau_i e^{-j\omega\tau_i}\right)u\right\}=0\\
\end{array}\right.}
\end{equation}
in  the $4n+2$ unknowns $\Re(v),\Im (v),\Re(u),\Im(u),\omega$ and
$\xi$. These equations are still overdetermined because  the
property (\ref{col1}) is not explicitly exploited in the
formulation, unlike the property (\ref{col2}). However, it makes the
equations (\ref{forfinal}) solvable in least
squares sense, and the $(\omega,\xi)$
components have a one-to-one-correspondence with the solutions of
(\ref{direct2x}).

In conclusion, as a result of the approximation step, the largest $\xi$ for which $\LxiN$ has the imaginary axis eigenvalues and their corresponding eigenvectors are the approximate results of the largest eigenvalue of $G$. Using these results as estimates of $(\hat\xi,\hat\omega)$ satisfying (\ref{direct0}) and $u$ and $v$, we can find the exact values by solving (\ref{forfinal}). At the end of the correction step, the exact $\Hi$ norm of $G$ (\ref{eq:tfG}) and the achieved frequency are equal to $\xi=\hat\xi$ and $\omega=\hat\omega$ respectively.

\section{Algorithm} \label{sec:Alg}

We present two algorithms which are based on the relations between the singular values of the transfer function $G(j\omega)$ and the spectrum of the operator $\mathcal{L}_{\xi}$, described in Theorem~\ref{thm:GLxi} and the correction method based on the nonlinear eigenvalue problem defined in (\ref{forfinal}). From these relations we
get:\vspace{-.25cm}
\begin{eqnarray}
\nonumber \|G(j\omega)\|_{\mathcal{H}_{\infty}}=\sup \{\xi\in\mathbb{R}_+:\
\mathrm{operator\ }\mathcal{L}_{\xi} \mathrm{\ has\ an} \\
\mathrm{eigenvalue\ on\ the\ imaginary\ axis} \}.
\end{eqnarray}
The fact that the infinite-dimensional operator $\mathcal{L}_{\xi}$ can be approximated with the matrix $\mathcal{L}_{\xi}^N$, as outlined in Section~\ref{sec:LxiN}, and the fact that an estimate of the $\Hi$ norm of $G$ can be corrected to the true
value, as outlined in Section~\ref{sec:HinfCor}, suggest the following predictor-corrector computational scheme:
\begin{enumerate}
\item  for fixed $N$, determine
\begin{eqnarray}\label{stap1}
\nonumber \sup \{\xi\in\mathbb{R}_+:\ \mathrm{matrix}\ \mathcal{L}_{\xi}^N
\mathrm{\ has\ an\ eigenvalue\ } \\
\mathrm{on\ the\ imaginary\ axis} \}
\end{eqnarray}
and determine the corresponding eigenvalues on the
imaginary axis;
\item  correct the results from the previous step by solving
the equations (\ref{forfinal}).
\end{enumerate}

Under a mild condition on the grid, the next theorem allows to interpret step $1$ as computing the $\Hi$ norm of an approximation of $G$.
\begin{theorem} \label{thm:Lxi-GN}
Assume that the mesh $\Omega_N$ is symmetric around the zero as given in (\ref{meshSymmetry}). Let $p_N$ be the polynomial of the degree $2N+1$ satisfying the conditions,\vspace{-.2cm}
\begin{eqnarray}
p_N(0;\ \lambda)&=&1, \\
p_N^{\prime}(\theta_i;\lambda)&=&\lambda p_N(\theta_i;\lambda),\;
\nonumber i=-N,\ldots,-1,1,\ldots,N.
\end{eqnarray}
Let $\xi>0$ be such that $\det(D^TD-\xi^2 I)\neq0$. The matrix $\LxiN$ has an imaginary axis eigenvalue $\lambda=j\omega$ if and only if $G_N(j\omega)$ has a singular value equal to $\xi$ where\vspace{-0.15cm}
\begin{eqnarray}\label{defGN}
{\textstyle
\nonumber G_N(j\omega)=C\left(j\omega I-A_0-\sum_{i=1}^m A_i p_N(-\tau_i;\
j\omega)\right)^{-1}B+D}.
\end{eqnarray}
\end{theorem}
\noindent\textbf{Proof.\ } The proof is given in Appendix, Section \ref{sec:App}.

\noindent\textbf{Remark:\ } As we shall see later, the functions $p_N(-\tau_i,\lambda)$ are proper rational functions in $\lambda$.

In what follows we assume that the grid $\Omega_N$, employed in the
discretization of $\mathcal{L}_{\xi}$, is symmetric around zero,\
i.e.\ it satisfies (\ref{meshSymmetry}). Theorem \ref{thm:Lxi-GN} guarantees that $\mathcal{L}_{\xi}^N$ has eigenvalues on the imaginary axis for all \vspace{-.2cm}
\[{\textstyle
\xi\in\left[\sigma_1(D),\
\|G_N(j\omega)\|_{\mathcal{H}_{\infty}}\right]}\vspace{-.2cm}
\]
and no eigenvalues on the imaginary axis for
\mbox{$\xi>\|G_N(j\omega)\|_{\mathcal{H}_{\infty}}$}. Thus the supremum in (\ref{stap1}) exists.

%

In the first algorithm, the prediction step
is based on the bisection algorithm presented
in~\cite{bisection:byers}.

\begin{algorithm}\label{algbis} \ \

{\small {\em
 \noindent Input: system data, $N$, symmetric grid $\Omega_N$,
 tolerance tol for prediction step\\
 Output: $\|G(j\omega)\|_{\mathcal{H}_{\infty}}$
 }
\begin{enumerate}
\item[]\hspace*{-0.6cm} \underline{Prediction step:}
\item compute a lower bound $\xi_l$ on $\|G_N(j\omega)\|_{\mathcal{H}_{\infty}}$,
e.g.\
$\xi_l:=\max\left\{\sigma_1(G(0)),\sigma_1(D),\mathrm{tol}\right\}$\\
set upper bound,\ $\xi_h:=\infty$
\item while $\xi_h-\xi_l>2\ \mathrm{tol}$
  \begin{itemize}
  \item[2.1] if $\xi_h=\infty $, set $\xi:= 2\xi_l$, else set
  $\xi=(\xi_l+\xi_h)/2$
  \item[2.2] compute $\mathcal{E}_{\xi}$, the set
  of eigenvalues of the matrix $\mathcal{L}_{\xi}^N$ on the positive imaginary
  axis
  \item[2.3] if $\mathcal{E}_{\xi}=\phi$, then $\xi_h=\xi$,  else $\xi_l=\xi$
  \end{itemize}
   \item[] \hspace*{-0.4cm}\{result: estimate $(\xi_h+\xi_l)/2$
 for $\|G_N(j\omega)\|_{\mathcal{H}_{\infty}}$\}
\item[]\hspace*{-0.6cm} \underline{Correction step:}
\item determine all eigenvalues
$\{j\omega^{(1)},\ldots,j\omega^{(p)}\}$ of
 $\mathcal{L}_{\xi_l}^N$ on the positive imaginary axis, and the corresponding
 eigenvectors $\left\{x^{(1)},\ldots,x^{(p)}\right\}$.
 \item for all $i\in\{1,\ldots,p\}$, solve (\ref{forfinal}) with
 starting values \vspace{-.3cm}
 \[
\left[\begin{array}{c}u\\ v\end{array}\right]=x^{(i)}_0,\
\omega=\omega^{(i)},\ \ \xi=\xi_l,\ \ \sigma=0
\vspace{-.2cm}
 \]
denote the solution with $(\tilde u^{(i)},\tilde
 v^{(i)},\tilde\omega^{(i)},\tilde\xi^{(i)})$.
\item set
$\|G(j\omega)\|_{\mathcal{H}_{\infty}}:=\max_{1\leq i\leq
p}\tilde\xi^{(i)}$.
\end{enumerate}
}
\end{algorithm}

\begin{figure}[!h]
\begin{minipage}[b]{0.49\linewidth}
\centering
\resizebox{4.7cm}{!}{\includegraphics{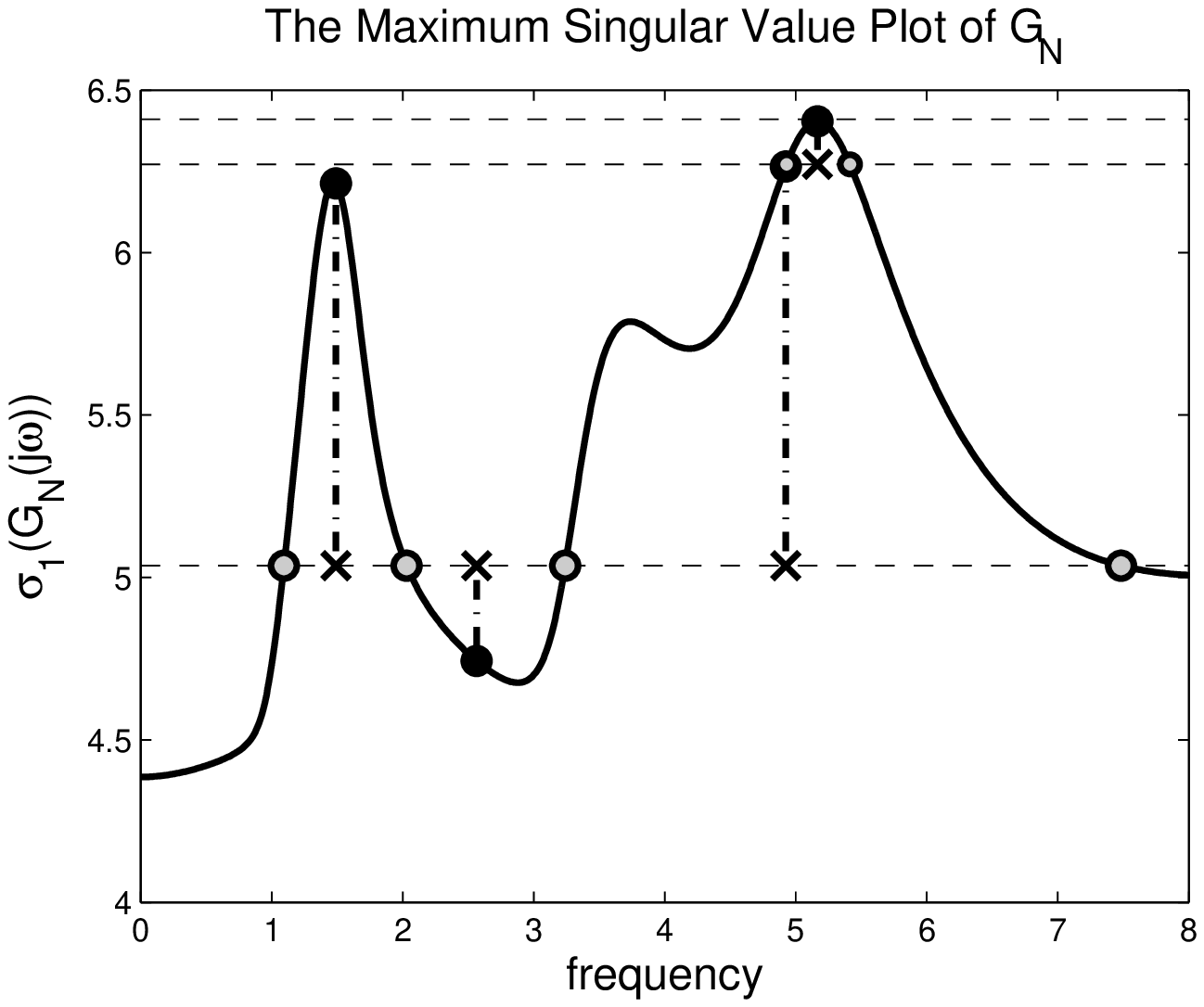}}
\caption{\label{fig:steinbuch} The maximum singular value plot of the function
$G_N(j\omega)$}
\end{minipage}
\begin{minipage}[b]{0.49\linewidth}
\centering
\resizebox{4.7cm}{!}{\includegraphics{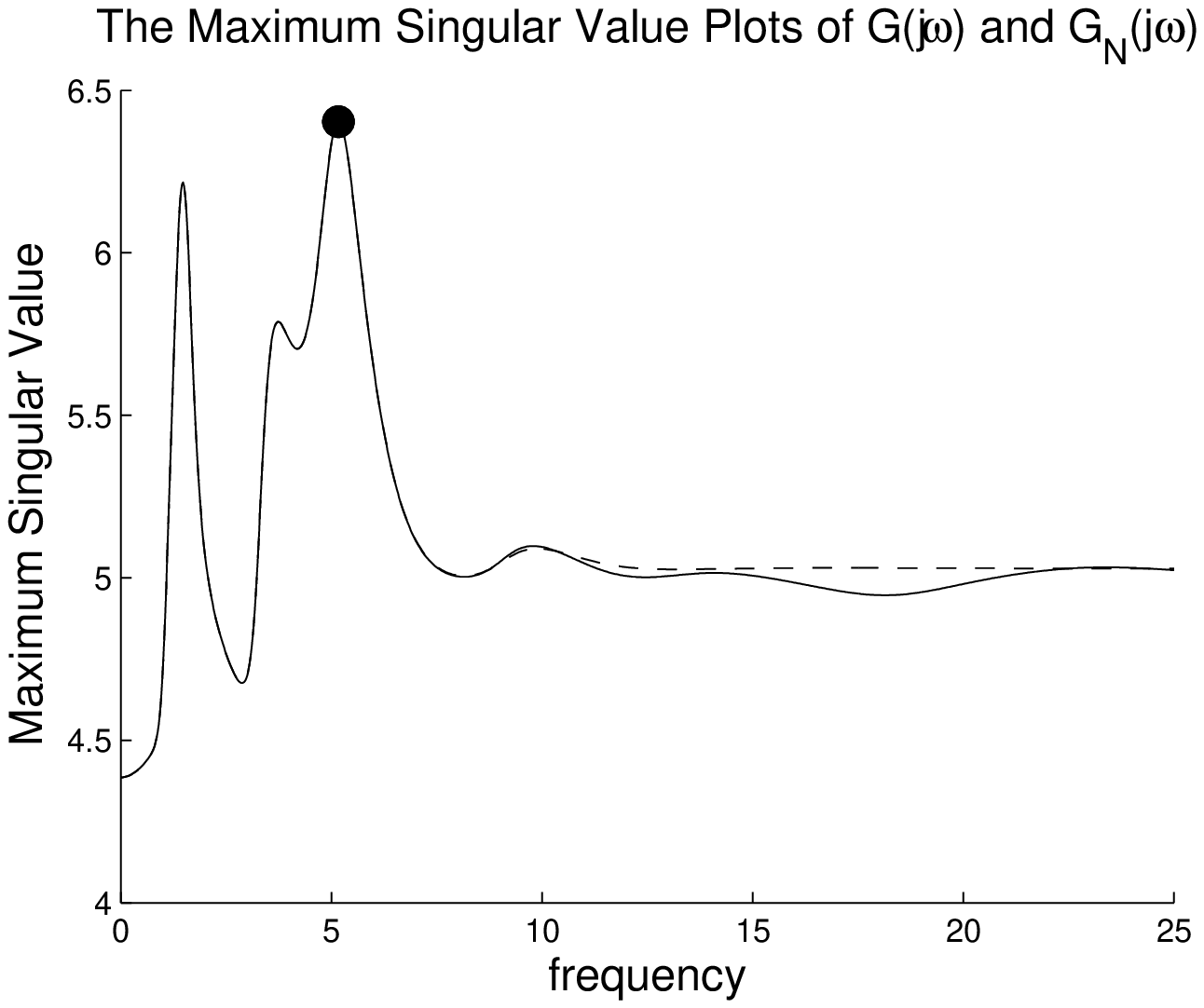}}
\caption{\label{fig:singularG} Singular value plot of the function
$G(j\omega)$}
\end{minipage}
\end{figure}

In the prediction step of the second algorithm, we use the fast iterative algorithm given in \cite{quad:bruinsma} for the prediction step: Given a transfer function $G_N$ defined in Theorem \ref{thm:Lxi-GN} and its corresponding \emph{Hamiltonian-like} matrix $\LxiN$ (by Theorem \ref{thm:Lxi-GN}), the largest singular values of $G_N$ are calculated as follows:
\begin{itemize}
  \item For a fixed level set $\xi$ (shown as dashed lines in Figure~\ref{fig:steinbuch}), calculate the imaginary axis eigenvalues of $\LxiN$ (shown in gray dots in Figure), these eigenvalues are also the frequencies of the singular values equal to $\xi$ by Theorem~\ref{thm:Lxi-GN},
  \item Find the middle points on each interval of the calculated frequencies (shown with cross signs in Figure), and calculate the largest singular value of each middle point (shown in black dots in the Figure),
  \item Set the next level set $\xi$ to the maximum of the calculated largest singular values at the middle points.
\end{itemize}

This algorithm \cite{quad:bruinsma} is quadratically convergent and well known method in the computation of $\Hi$ norms for the finite dimensional systems. The overall algorithm for the computation of $\Hi$ norm of (\ref{eq:tfG}) becomes:

\begin{algorithm}\label{algmiddle} \ \

{\small
 {\em
 \noindent Input: system data, $N$, symmetric grid $\Omega_N$, candidate
critical frequency $\omega_t$ if available,\\
\hspace*{1cm}tolerance tol for prediction step\\
 Output: $\|G(j\omega)\|_{\mathcal{H}_{\infty}}$
 }
\begin{enumerate}
\item[]\hspace*{-0.6cm} \underline{Prediction step:}
\item compute a lower bound $\xi_l$ on
$\|G_N(j\omega)\|_{\mathcal{H}_{\infty}}$,\\
e.g.\
$\xi_l:=\max\left\{\sigma_1(G(0)),\sigma_1(D),\mathrm{tol}
, \sigma_1(G_N(j\omega_t))\right\}$
\item repeat until break
  \begin{itemize}
  \item[2.1] set  $\xi:=\xi_l(1+2\ \mathrm{tol})$
  \item[2.2] compute the set
  of eigenvalues $\mathcal{E}_{\xi}$ of the matrix $\mathcal{L}_{\xi}^N$ on the positive imaginary
  axis,\\
   $\mathcal{E}_{\xi}:=\left\{j\omega^{(1)},j\omega^{(2)},\ldots
   \right\}$, with $0\leq \omega^{(1)}<\omega^{(2)}<\ldots$
  \item[2.3] if $\mathcal{E}_{\xi}=\phi$, break\\
  else\\
  $\ \ \ \mu^{(i)}:=\sqrt{\omega^{(i)}\omega^{(i+1)}},\
  i=1,2,\ldots$\\
  compute $\sigma_1(G_N(j \mu^{(i)})),\ i=1,2,\ldots$\\
  set $\xi_l:= \max_i\sigma_1(G_N(j \mu^{(i)}))$
  \end{itemize}
 \item[] \hspace*{-0.4cm}\{result: estimate $(\xi+\xi_l)/2$
 for $\|G_N(j\omega)\|_{\mathcal{H}_{\infty}}$\}
\item[]\hspace*{-0.6cm} \underline{Correction step:}\\
\hspace*{-0.4cm}follow the steps 3.-5. of
Algorithm~\ref{algbis}
\end{enumerate}
}
\end{algorithm}

In Step 2.3 of Algorithm \ref{algmiddle}, we need the evaluation of the $G_N(j\omega)$ at specific frequencies. This can be done as follows:
\textbf{Evaluation of $G_N$} \\
Algorithm~\ref{algmiddle} relies on the evaluation of the
function $G_N$, and, hence, on the evaluation of the
polynomials $p_N(-\tau_i;\ \lambda),\ i=1,\ldots,m$ for
several values of $\lambda$.
Given the polynomial basis $B_i(t)$, we represent $p_N(\cdot;\
\lambda)$:\vspace{-.2cm}
\[{\textstyle
p_N(t;\ \lambda)=\sum_{i=0}^{2N}\alpha_i
B_i(t)}.
\vspace{-.2cm}
\]
 From its definition $p_N(\cdot;\
\lambda)$ satisfies the conditions
\begin{eqnarray}
\nonumber && \hspace{-1.9cm} p_N(0;\ \lambda)=1,\ \textrm{and} \ p_N^{\prime}(\theta_i;\
\lambda)=\lambda p_N(\theta_i;\ \lambda),\ \\
&&  \hspace{1cm} i\in\{-N,\ldots,N\}\cup\{1,\ldots,N\}.
\end{eqnarray}\vspace{-.2cm}
For $\lambda\neq 0$, the conditions can be written as
\begin{equation}\label{solvepn}
{\textstyle
\left(\lambda \left[\begin{array}{l}
0_{1\times(2N+1)} \\
M
\end{array}\right]-\left[\begin{array}{l}
b \\
N
\end{array}\right]
\right)\alpha =\left[\begin{array}{l} -1 \\0_{2N\times1} \end{array}\right]},
\end{equation}
where $M_{ij}=B_{j-1}(\theta_{i-(N-1)})$, $N_{ij}={B'}_{j-1}(\theta_{i-(N-1)})$, $b_{1j}=B_{j-1}(0)$ and $\alpha_{j1}=\alpha_{j-1}$ for $i=1,\ldots,2N$ and \mbox{$j=1,\ldots, 2N+1$}.

After solving (\ref{solvepn}) for a given value of
$\lambda$ we can evaluate\vspace{-.2cm}
\[{\textstyle
p_N(-\tau_i;\ \lambda)=\sum_{i=0}^{2N} \alpha_i
B_i(-\tau_i),\ \
i=0,\ldots,m}.
\]
\begin{remark} Although the prediction step in
Algorithm~\ref{algmiddle} corresponds to computing
$\|G_N(\lambda)\|_{\mathcal{H}_{\infty}}$, the matrix function
$G_N(\lambda)$  or the rational functions $p_N(\tau_i;\ \lambda)$
never need to be explicitly computed (note that they stem from a
particular \emph{interpretation} of the effect of a spectral
discretization of the operator $\mathcal{L}_{\xi}$ into the matrix
$\mathcal{L}_{\xi}^N$). Algorithm~\ref{algmiddle} only relies on
computing the eigenvalues of $\mathcal{L}_{\xi}^N$ and on evaluating
$G_N(j\omega)$ at \emph{specific frequencies}.
\end{remark}
\begin{remark}
The definition of $G_N$ in (\ref{defGN}) interprets the term $p_N(t,\lambda)$ as an approximation of the term $e^{\lambda t}$ over the whole interval $[-\tau_{\max},\tau_{\max}]$ where \mbox{$\tau_{max}=\max\{\tau_1,\ldots,\tau_m\}$}. Note that the use of the well-known Pad\'{e} approximation for the time- delay will cause numerically bad-scaled matrix in $\mathcal{L}_\xi^N$ due to the different magnitudes in the Pad\'{e} coefficients. Note that the Pad\'{e} approximation depends on the time-delay and for multiple delays, each delay is approximated separately which will increase the $\mathcal{L}_\xi^N$ dimension considerably. However, the term $p_N(t,\lambda)$ approximates multiple delays with a single term.
\end{remark}
\begin{remark}
Note that the prediction and correction steps are to some extent independent of each other.
In particular, other choices for a finite-dimensional approximation in the prediction
step are possible (e.g., using Pad\'{e}-like approximations or the frequency grid).
\end{remark}
\begin{remark}
The numerical method for computing $\Hi$ norm can be used for computing $\mathcal{L}_\infty$ norm of the time-delay system without any modification.
\end{remark}
\vspace{-.2cm}
\section{Example} \label{sec:Ex}
The time-delay system (\ref{eq:tfG}) has the dimensions as $m=7$, $n=10$, $n_u=2$, $n_y=4$ with delays $\tau_1=0.1$, $\tau_2=0.2$, $\tau_3=0.3$, $\tau_4=0.4$, $\tau_5=0.5$, $\tau_6=0.6$, $\tau_7=0.8$.

To illustrate insights of the algorithm and results, the maximum singular value plot of the transfer function $G(j\omega)$ (\ref{eq:tfG}) and that of the discretized transfer function $G_N(j\omega)$ are shown in Figure~\ref{fig:singularG} with blue and red lines where \mbox{$N=6$}. Note that the approximated transfer function has almost same behavior until $\omega=10$. The iterations in the prediction step of the second algorithm can be seen in Figure \ref{fig:steinbuch}. After three level set iterations, the prediction step yields $\xi=6.0436$ and the frequencies $\omega^{(1)}=5.1660$ and $\omega^{(2)}=5.1666$. Two frequencies converge to the peak of the maximum singular value plot $\xi=6.4040$ at $\tilde{\omega}^{(1)}=\tilde{\omega}^{(2)}=5.1662$. Therefore, the $\Hi$ norm of the time-delay system is $\|G(j\omega)\|_{\Hi} = 6.4040$.

The problem data for the above benchmark example and a MATLAB implementation of our code for the $\Hi$ norm computation are available at the website
{\small \verb"http://www.cs.kuleuven.be/~wimm/software/hinf"}.
\section{Conclusion} \label{sec:Conc}
A numerically stable method to compute $\Hi$ norm of time-delay system with arbitrary number of delays is
given. As a generalization of the finite dimensional case, we show the connection between singular values of a transfer function and the
eigenvalues of an infinite dimensional linear operator, equivalent to the Hamiltonian matrix in delay free case. By the discretization of the infinite dimensional linear operator, an approximation of $\Hi$ norm of the time-delay system is found. This result is corrected using the equations based on the nonlinear eigenvalue problem. The algorithms are easily extendable to the systems with distributed delays.
\vspace{-.2cm}
\section{Acknowledgement}
This article present results of the Belgian Programme on
Interuniversity Poles of Attraction, initiated by the Belgian State,
Prime Minister's Office for Science, Technology and Culture, and of
OPTEC, the Optimization in Engineering Center of the K.U.Leuven.

\vspace{-.2cm}
\section{Appendix} \label{sec:App}
\noindent\textbf{Proof of Theorem \ref{thm:Lxi-Hxi}.\ }
Assume that $\mathcal{L}_{\xi}\ u=\lambda
u$ holds. By (\ref{def:Lxi}), we obtain $u(t)=e^{\lambda t} v,\
t\in[-\tau_{\max},\ \tau_{\max}]$, with $v\in\mathbb{C}^{2n}$.
Taking into account the boundary condition (\ref{def:Lxi2}), the
nonlinear eigenvalue problem is satisfied, $H_{\xi}(\lambda)v=0$.
Conversely, if $H_{\xi}(\lambda)v=0$, then it is readily verified
that $u\equiv ve^{\lambda\theta},\ \theta\in[-\tau_{\max},\
\tau_{\max}]$, belongs to $\mathcal{D}(\mathcal{L}_{\xi})$ and
satisfies $(\mathcal{L}_{\xi}-\lambda I)u=0$.~$\Box$

The following theorem shows the connection between the singular value of $G$ equal to $\xi$ and the imaginary axis eigenvalue of the nonlinear eigenvalue problem (\ref{prob:HxiEig}).

\begin{theorem} \label{thm:Hxi}
Let $\xi> 0$ be such that the matrix \mbox{$\det(D_\xi)\neq 0$}. For $\omega\geq 0$, the matrix $G(j\omega)$ has a singular value equal to $\xi$ if and only if $\lambda=j\omega$ is a solution of the equation\vspace{-.2cm}
\begin{equation} \label{prob:nonlinear-eigenvalue}
\det H_{\xi}(\lambda)=0,
\vspace{-.2cm}
\end{equation}
where $D_\xi$ and $H_{\xi}$ are defined in Theorem \ref{thm:Lxi-Hxi}.
\end{theorem}

\noindent\textbf{Proof.\ } The proof is similar to the proof of Proposition~22 in \cite{genin}. For all $\omega\in\mathbb{R}$, we have the relation\vspace{-.2cm}
\begin{multline}\label{vier}
\det H_{\xi}(j\omega)\det D_{\xi}(j\omega)= \det
(G^*(j\omega)G(j\omega)-\xi^2
I)\\
\det\left(\left[\begin{array}{cc}A(j\omega) &0\\
0& -A(j\omega)^{*}
\end{array}\right]\right),
\end{multline}
where $A(j\omega)=j\omega I- A_0-\sum_{i=1}^m A_i
e^{-j\omega\tau_i}$. Both left and right hand side can be
interpreted as expressions for the determinant of the 2-by-2 block
matrix\vspace{-.1cm}
\[{\textstyle
\left[\begin{array}{cc|c} A(j\omega) &0& -B\\
C^TC&-A(j\omega)^{*}&C^TD \\
\hline D^T C &B^T&D_{\xi}
\end{array}\right]}
\vspace{-.1cm}
\]
using Schur complements. Since $D_{\xi}$ is non-singular and $G$ is
stable, we get from (\ref{vier}): \vspace{-.2cm}
\[{\textstyle
\det (G^*(j\omega)G(j\omega)-\xi^2 I)=0\Leftrightarrow\det
H_{\xi}(j\omega)=0.}
\vspace{-.2cm}
\]
This is equivalent to the assertion of the theorem.
 \hfill $\Box$

\noindent\textbf{Proof of Theorem \ref{thm:GLxi}:} Theorem \ref{thm:GLxi} follows from Theorem \ref{thm:Hxi} and  Theorem \ref{thm:Lxi-Hxi}.

\noindent\textbf{Proof of Proposition \ref{prop:LxiSym}:} It can be verified
that\vspace{-.2cm}
\[{\scriptstyle
H_{\xi}(-\bar\lambda)=-\left( \left(
\left[\begin{array}{rr}0&-1\\1&0
\end{array}\right]\otimes I\right)H_{\xi}(\lambda)
\left(\left[\begin{array}{rr}0&1\\-1&0
\end{array}\right]\otimes I\right)\right)^*},
\vspace{-.2cm}
\]
hence,\vspace{-.2cm}
\begin{equation}\label{symmetrie}
{\textstyle \det H_{\xi}(-\bar\lambda)=\left(\det
H_{\xi}(\lambda)\right)^*.}
\end{equation}
By Theorem \ref{thm:Lxi-Hxi}, the proposition follows. \hfill $\Box$

\noindent\textbf{Proof of Theorem \ref{thm:Lxi-GN}:} As in the continuous case the discretized linear eigenvalue problem\vspace{-.2cm}
\begin{equation}\label{infd}
{\textstyle \mathcal{L}_{\xi}^N\ x=\lambda x,\ \lambda\in\mathbb{C},\
x\in\mathbb{C}^{(2N+1)2n},\ x\neq 0,}
\vspace{-.2cm}
\end{equation}
has a nonlinear eigenvalue problem of dimension $2n$ as counterpart. To see this, we get from (\ref{defldisc}) and (\ref{infd}):\vspace{-.2cm}
\begin{eqnarray}
&& \nonumber \hspace{-1cm} {\textstyle (\mathcal{P}_Nx)^{\prime}(\theta_{N,i})=\lambda x_i=\lambda \mathcal{P}_Nx(\theta_{N,i})} \\
&& \hspace{2.5cm} \textrm{for} \ \ {\textstyle i\in\{-N,\ldots,N\},\ i\neq0}  \label{een} \\
&& \nonumber \hspace{-1.0cm} M_0 \mathcal{P}_N x(0)+\sum_{i=1}^m
\left(M_i \mathcal{P}_N
x(-\tau_i)+ M_{-i} \mathcal{P}_N x(\tau_i)\right)=\lambda x_0 \\
&& \hspace{5cm}=\lambda\mathcal{P}_N x(0). \label{twee}
\vspace{-.2cm}
\end{eqnarray}
From  $\mathcal{P}_N x(0)=x_0$ and (\ref{een}) it follows that\vspace{-.2cm}
\begin{equation}\label{collocation}
\mathcal{P}_N x(\cdot)=p_N(\cdot;\ \lambda) x_0,
\vspace{-.2cm}
\end{equation}
where $p_N(\cdot;\ \lambda):\ \mathbb{R}\to\mathbb{C}$ is the
\emph{collocation polynomial} for the equation\vspace{-.2cm}
\begin{equation}\label{col}
{\textstyle \dot z(t)=\lambda z(t),\ \ z,\lambda\in\mathbb{C},}
\vspace{-.2cm}
\end{equation}v
which satisfies (\ref{col}) on $\Omega_N\setminus\{0\}$, as well the
interpolating condition $p_N(0;\ \lambda)=1$.
Note that for a fixed value of $t$ the function $p_N(t;\ \lambda)$
is a rational function in $\lambda$.
When substituting (\ref{collocation}) in (\ref{twee}) we arrive at
the discretized nonlinear eigenvalue problem (\ref{nonlineardiscrete}) and (\ref{defhn}),\vspace{-.2cm}
\begin{equation}\label{nonlineardiscrete}
{\textstyle H^N_{\xi}(\lambda)\ x_0=0,}
\vspace{-.2cm}
\end{equation}
where\vspace{-.2cm}
\begin{equation}\label{defhn}
{\scriptstyle H_{\xi}^N(\lambda)=\lambda
I-M_0-\sum_{i=1}^m \left(M_i p_N(-\tau_i;\
\lambda)+M_{-i}p_N(\tau_i,\ \lambda)\right).}
\vspace{-.2cm}
\end{equation}
and the matrices $M_0$, $M_i$, $M_{-i}$ are defined in Theorem~\ref{thm:GLxi}.
The nonlinear eigenvalue problem (\ref{nonlineardiscrete}) is equivalent to the linear infinite dimensional eigenvalue problem (\ref{infd}). The expressions (\ref{nonlineardiscrete})-(\ref{defhn}) can also be interpreted as a direct approximation of (\ref{prob:HxiEig}) and (\ref{eq:HamMatrix}).

By Proposition \ref{propsymLxi}, the eigenvalues of (\ref{infd}) are symmetric with respect to the imaginary axis. Using the equivalence of (\ref{infd}) and (\ref{nonlineardiscrete}), same symmetry property is valid for (\ref{nonlineardiscrete}). The assertion follows from the arguments mentioned in the proof of Theorem \ref{thm:Hxi}. \hfill $\Box$

\noindent\textbf{Proof of Proposition \ref{propsymLxi}:}
The condition on the mesh assures that\vspace{-.3cm}
\begin{equation}\label{propp}
{\textstyle p_N(-\tau_i;\ \lambda)=p_N(\tau_i;\ -\lambda),\
\forall\lambda\in\mathbb{C},\ \forall i=0,\ldots,N.}
\vspace{-.2cm}
\end{equation}
Next, using the same arguments as in the proof of
Proposition~\ref{prop:LxiSym} we arrive at\vspace{-.2cm}
\begin{equation} \label{eq:HNSym}
{\textstyle \det H^N_{\xi}(-\bar\lambda)=(\det H^N_{\xi}(\lambda))^*.}
\vspace{-.2cm}
\end{equation}
The Proposition follows from (\ref{eq:HNSym}) and the arguments mentioned in the proof of Theorem \ref{thm:Lxi-GN} on the equivalence of the eigenvalue problems (\ref{infd}) and (\ref{nonlineardiscrete}). \hfill $\Box$
\end{document}